\documentclass[10pt]{article}
\usepackage[T1]{fontenc}
\PassOptionsToPackage{authoryear,round}{natbib}
\usepackage[preprint]{neurips_2026}
\renewcommand{\acksection}{\section*{Acknowledgement}}
\usepackage{graphicx,booktabs,amsmath,amssymb}
\usepackage{float}
\usepackage{url}
\usepackage{xurl}
\usepackage{microtype}
\usepackage{hyperref}
\hypersetup{hidelinks,pdftitle={Better Behavioral Prediction, More Faithful Model Ablations? Evidence from Sequential Choice},pdfauthor={Hanbo Xie}}
\graphicspath{{figures/}}
\title{Better Behavioral Prediction, More Faithful Model Ablations? Evidence from Sequential Choice}
\author{Hanbo Xie\\
School of Psychological and Brain Sciences\\
Georgia Institute of Technology\\
Atlanta, GA 30332, USA\\
\texttt{hanboxie1997@gatech.edu}}
\newcommand{\nll}{\operatorname{NLL}}
\newcommand{\err}{\mathcal{E}}
\newcommand{\softmax}{\operatorname{softmax}}
\begin{document}
\maketitle

\begin{abstract}
Using predictive models to explain cognition requires more than accurate behavioral predictions. Input ablations offer an appealing route: remove information from a model and interpret the resulting performance change as evidence of its importance for behavior. Yet this inference assumes that the model's dependence on information reflects the dependence of the process generating the behavior. We test it in two synthetic sequential bandit tasks with known generating policies, where past choices can remain informative when feedback is unavailable to a predictor. We compare GRUs and Transformers trained from scratch, a fine-tuned LLaMA model, and cognitive models across systematically varied reward contributions. Our analyses distinguish prediction after training without reward observations from the response of a fixed predictor to donor-reward replacement. Three findings emerge. First, in the restless task, neural models trained without rewards predict held-out choices better than four simple training-fitted behavioral baselines. Second, under matched donor replacement, accurate predictors can respond much less than the known generator. Third, at some reward weights, neural networks predict better than a pooled reinforcement-learning model but have less faithful changes in choice probabilities; the model ordering differs between the two tasks. These independent-test results separate information sufficient for prediction from response fidelity under a specified ablation in sequential choice. They motivate validating model-ablation responses independently of predictive performance before using them to infer how the observed behavior was generated.
\end{abstract}

\section{Introduction}
Predictive models of human behavior are becoming more expressive. Recurrent neural networks capture sequential choice patterns that conventional cognitive models can miss, including with very small networks whose dynamics can be inspected \citep{dezfouli2019models,ji2025discovering}. Large behavioral datasets and pretrained language models further extend the scale and range of behavioral prediction \citep{peterson2021using,binz2025foundation}. The scientific appeal extends beyond forecasting: an accurate behavioral model might also serve as a proxy for studying the processes that produce behavior. For that explanatory use, the model's response to a proposed test needs validation in addition to its choice predictions.

\begin{figure}[p]
\centering
\includegraphics[width=\linewidth]{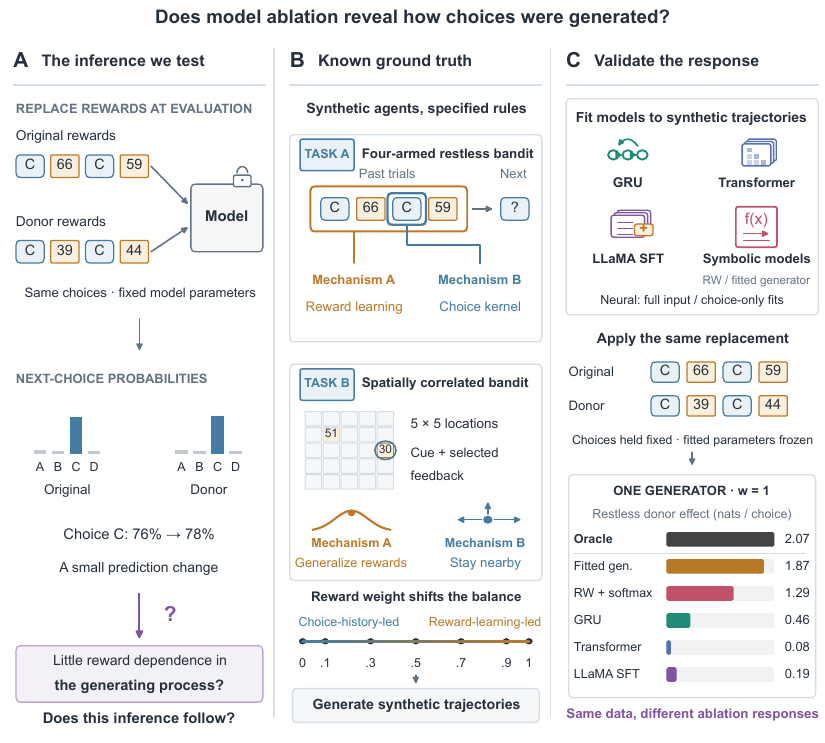}
\caption{\textbf{Reward-history ablation and claims about sequential behavior.} \textbf{(A)} At evaluation, past rewards are replaced while observed choices and fitted model parameters remain fixed. The generic model diagram shows how a small prediction change can invite an inference of weak reward dependence in the process generating behavior. The arrow and question mark mark the inference being tested. \textbf{(B)} Task A (four-armed restless bandit) combines reward learning with a last-choice kernel. Task B (spatially correlated bandit) combines reward generalization with local choice tendencies. Seven weights control the relative contributions to the known generating policies. \textbf{(C)} Neural and symbolic predictors are fitted to synthetic trajectories and evaluated under matched original and donor-replaced reward histories. Neural models have separate full-input and choice-only fits; symbolic predictors comprise RW + softmax and the fitted generator. The oracle uses true generating parameters. Bars show mean test-set increases in negative log likelihood at reward weight one in Task A over 20 shared donor assignments. Neural models and RW use pooled training; fitted generators use individual prefix calibration (Section~\ref{sec:methods}). A retains a validation-trial Transformer example to illustrate the operation; C reports aggregate test responses. Blue marks choices; gold marks rewards. Oracle replay conditions on observed choices throughout.}
\label{fig:design}
\end{figure}

In a reward-learning task, input ablation provides a seemingly direct test: researchers remove or replace reward information in a fitted model's input and measure the change in its predictions. A small effect can invite the claim that, because the model predicts almost as well without rewards, the people whose choices it was trained to predict must also have relied little on rewards. This moves from an observation about the fitted model to an inference about how the observed human choices were generated. The inference need not hold. In sequential tasks, previous choices are themselves consequences of earlier rewards and persistent individual tendencies. They can therefore support accurate next-choice prediction even when rewards are not directly available. Conversely, a model's response to reward perturbation can depend on its representation, training distribution, and the perturbation procedure.

These concerns are familiar in machine learning. Work on removal-based explanations distinguishes different ways of removing information and different quantities being explained \citep{covert2021explaining}. Retraining controls address the distribution shift introduced by feature removal \citep{NEURIPS2019_fe4b8556}, and explanation sanity checks show why explanatory outputs require their own validation \citep{adebayo2018sanity}. In behavioral modeling, concerns about sequential shortcuts have also arisen in evaluations of Centaur \citep{xie2025centaur}. We ask a quantitative question for sequential choice: \emph{how closely does a behavioral predictor's response to reward-history replacement match the response of the process that generated its data under the same operation?}

We address this question in two synthetic bandit tasks with known generators and systematically varied reward contributions. We compare GRUs, Transformers, and fine-tuned LLaMA with cognitive references, separating training without rewards from reward replacement in a fixed predictor. Choice-only neural models retain nontrivial predictive information; accurate full-input neural models can nevertheless respond much less than the generator to matched reward replacement. Moreover, predictive and response-fidelity rankings differ between the two tasks, without uniformly favoring neural or cognitive models. In these settings, ablation magnitude is a property of the fitted model and chosen operation, not a direct estimate of the generating weight.

\section{Related work}
\paragraph{Predictive models as cognitive instruments.}
Behavior-trained RNNs offer flexible alternatives to hand-specified learning rules \citep{dezfouli2019models}; large-scale behavioral modeling can also guide the development of interpretable theories \citep{peterson2021using}. Centaur extends behavioral prediction through language-model fine-tuning on heterogeneous cognitive experiments \citep{binz2025foundation}. Our LLaMA comparison uses the pretrained base model, not Centaur's released adapter or its multi-task training. The choice-history concern raised by \citet{xie2025centaur} motivates our first comparison. Recent prompt-channel and trial-order diagnostics find task-dependent information use in cognitively fine-tuned language models \citep{oh2026smallcogfm}. These evaluation-time removals differ from our choice-only retraining. The authors also distinguish prediction from process explanation; our known generators quantify whether the fitted model's response matches the generating policy under the same operation.

\paragraph{From prediction to mechanism evidence.}
Neural behavioral models can be interrogated beyond held-out accuracy. \citet{dezfouli2019models} use off-policy simulations to characterize learned decision strategies. The tinyRNN framework analyzes low-dimensional network dynamics to make learned updates inspectable \citep{ji2025discovering}. In a four-armed drifting bandit, \citet{eckstein2026hybrid} combine cognitive and neural modules, probe what the fitted modules encode, and test whether model-generated choices reproduce human behavioral patterns. These approaches provide different kinds of evidence for a mechanistic interpretation. We ask a complementary, narrower question: when rewards are replaced in the input to a fixed predictor, does its change in choice probabilities match the change in the process that generated its data? Our known-generator experiments calibrate that observable response, rather than adjudicating the internal-mechanism claims of these studies.

\paragraph{Removal, retraining, and explanation validation.}
Removal-based explanations depend on how features are removed, what model behavior is measured, and how individual effects are summarized \citep{covert2021explaining}. ROAR retrains after feature removal to separate retained predictive information from degradation in an unadapted model \citep{NEURIPS2019_fe4b8556}. Our choice-only training has a related purpose, but it does not validate an evaluation-time ablation by itself. Sanity checks \citep{adebayo2018sanity} and shortcut-learning research \citep{geirhos2020shortcut} further caution against identifying predictive success with the intended computation. Simulation, parameter recovery, and model recovery are established checks for specified cognitive models \citep{wilson2019ten}. We apply an analogous validation principle to neural-model ablations by comparing their probability responses with a known generating process under the same input operation. We do not claim that input ablation recovers a neural circuit or uniquely identifies an internal algorithm.

\paragraph{Reward learning and choice history.}
Reward prediction and value updating are central to computational accounts of sequential choice \citep{schultz1997neural,daw2006computational}. Yet the act of choosing can also influence later choices independently of its outcome: \citet{lau2005dynamic} found that past rewards and past choices jointly described trial-by-trial behavior. This distinction echoes the long-standing contrast between the law of effect, which links repetition to reward, and the law of exercise, which links it to performing an action \citep{gershman2020origin}. A repeated choice can therefore reflect updated reward expectations, an outcome-independent choice tendency, or both. We vary these two influences in controlled generators, using a signed last-choice term in the restless task and a local-choice tendency in the spatial task. These components are simplified test cases, not proposed accounts of human cognition.

\section{Ablation responses as a separately validated target}
\label{sec:estimands}
Let $a_{it}$ and $r_{it}$ denote participant $i$'s choice and chosen reward on trial $t$. The available history $h_{it}$ contains task information and observations strictly preceding the target choice. A fitted predictor returns $p_\theta(a\mid h_{it})$. The generating policy, evaluated with participant-specific true parameters, returns $o_i(a\mid h_{it})$. Every reported probability is normalized over the task's valid choices. Models are evaluated by mean choice negative log likelihood (NLL, in nats), with participants weighted equally.

\paragraph{Training without rewards.}
We train separate predictors on full histories and on histories with reward observations omitted. Their difference,
\begin{equation}
G=\nll(p_{\rm choice})-\nll(p_{\rm full}),
\label{eq:gain}
\end{equation}
measures the predictive benefit of reward access for these fitted pipelines. It is not the causal effect of reward on the generating agent, nor a guarantee that either predictor reaches the optimal conditional distribution. Strong choice-only performance establishes retained predictive information; it does not identify whether that information reflects inferred reward structure or other behavioral regularities.

\paragraph{Intervening on a fixed predictor.}
Our primary operation is participant-wise reward-sequence permutation, not literal removal of reward tokens. We use it to align model and oracle inputs. Appendix~\ref{app:operations} contrasts its temporal scope and reports earlier deletion and placeholder controls; these operations are not interchangeable definitions of ablation.
For evaluation-time ablation, we randomly permute participants within the same weight condition, with no participant assigned to themselves. This one-to-one mapping $\pi$ is a \emph{donor assignment}: participant $i$ receives participant $\pi(i)$'s recorded chosen-reward sequence. For example, if participant $j$ is assigned to $i$, predicting $i$'s choice on trial $t$ uses $i$'s actual choices on trials $1{:}t-1$ but $j$'s rewards at those trial indices. The donor stays fixed across the sequence; we do not independently shuffle rewards at each trial or replace any choices. In the spatial task, map and within-map trial indices are matched, and the recipient's initial cue is unchanged. Denote this history $D_\pi h$. We use 20 fixed assignments shared across model families. The scalar effect is
\begin{equation}
\Delta\nll(p)=\mathbb{E}_{i,t,\pi}\!\left[-\log p(a_{it}\mid D_\pi h_{it})+\log p(a_{it}\mid h_{it})\right].
\label{eq:delta}
\end{equation}
This intervention is a reward replacement, not literal deletion. It maintains reward observations while disrupting their correspondence with the recipient's history. All previous and target choices remain those in the original dataset. Thus the assay is teacher-forced, one-step-ahead evaluation, not a new closed-loop rollout in an altered environment.

\paragraph{Calibrating the response.}
The oracle receives exactly the same histories and replacement reward sequences as each fitted predictor. Define the probability-response vectors $v_p=p(\cdot\mid D_\pi h)-p(\cdot\mid h)$ and $v_o=o(\cdot\mid D_\pi h)-o(\cdot\mid h)$. Our response-fidelity measure is
\begin{equation}
\err(p,o)=\mathbb{E}_{i,t,\pi}\left[\frac{1}{2}\sum_a |v_p(a)-v_o(a)|\right].
\label{eq:error}
\end{equation}
Lower values indicate closer changes in choice probabilities. Unlike a difference in aggregate NLL effects, this measure does not let errors on different actions or trials cancel before comparison. It compares signed response vectors, so it is not a total-variation distance between two probability distributions and is not restricted to one. The oracle has zero error by construction. Matching this observable response is narrower than recovering the generator's latent states or unique implementation.

Equations~\ref{eq:gain}--\ref{eq:error} answer distinct questions. We do not divide model effects by oracle effects to report a percentage of mechanism recovery: the oracle effect can be small, and its dependence on the generating weight also reflects the policy and induced histories. Likewise, $G$ and $\Delta\nll$ do not form an additive decomposition of information. For this assay, we therefore report original NLL, perturbed NLL, their difference, and probability-response error.

\section{Experimental design}
\label{sec:methods}
\paragraph{Two generators and seven weights.}
Both tasks combine learning about rewards with choice tendencies that do not directly use rewards. Separating reward and choice history has precedent in sequential behavioral modeling \citep{lau2005dynamic}; delta-rule learning supplies a conventional reward-learning component \citep{dezfouli2019models}. Our specific combinations are controlled synthetic generators, not fitted accounts of human behavior. Both use $w\in\{0,.1,.3,.5,.7,.9,1\}$, with separate training at each weight. This design tests models of reward and choice histories; it does not test feature ablations in independent-trial decisions.

In the \textbf{restless task}, four independently evolving reward means produce a 200-trial session. Only the chosen value is updated, and the policy combines normalized values with a signed last-choice indicator:
\begin{align}
Q_{i,t+1,a_{it}}&=Q_{it,a_{it}}+\alpha_i(r_{it}-Q_{it,a_{it}}),\qquad K_{it,a}=\mathbf{1}[a=a_{i,t-1}],\\
u_{it,a}&=w\frac{Q_{it,a}-\bar Q_{it}}{c_Q}+(1-w)s_iK_{it,a},\qquad
o_i(a\mid h_{it})=.9\frac{\mathbf{1}[a\in M_{it}]}{|M_{it}|}+.1/4,
\label{eq:restlesspolicy}
\end{align}
where $M_{it}=\arg\max_a u_{it,a}$, $\alpha_i$ is a learning rate, and $c_Q$ is a fixed reward scale. The sign $s_i=+1$ promotes repetition and $s_i=-1$ penalizes the last choice; $K=0$ on the first trial. This one-step kernel is not a gradually accumulated trace. Increasing $w$ changes the relative utility contributions before an $\epsilon$-greedy choice, not a linear mixture of action probabilities.

In the \textbf{spatial task}, each participant encounters eight independent $5\times5$ reward maps, with 20 choices per map and one initially revealed location/reward. Following Gaussian-process approaches to spatial reward generalization \citep{wu2018generalization}, a learner updates posterior means $m_{it}(a)$ from observed rewards. We add a reward-independent local-choice component:
\begin{equation}
o_i(a\mid h_{it})=.9\left[w\softmax(m_{it}/\tau_{R,i})_a
+(1-w)\softmax(-d_1(\cdot,a_{i,t-1})/\tau_{L,i})_a\right]+.1/25.
\label{eq:spatialpolicy}
\end{equation}
Here $d_1$ is Manhattan distance from the last visited position, and $\tau_R,\tau_L$ control the two components' concentration. Unlike Eq.~\ref{eq:restlesspolicy}, $w$ mixes action probabilities directly. Values and histories reset between maps. Individual parameters vary within conditions. Appendix~\ref{app:generators} gives environmental distributions; Appendix~\ref{app:generatorchecks} reports behavioral summaries, existing parameter fits, and a spatial component-model recovery pilot, including identifiability limits.

The restless task has 1,000 training, 250 validation, and 250 test participants per weight; the spatial task has 200, 50, and 50, respectively. Participants have independent environmental streams, not a shared schedule library. Across weights, the same base participants share parameters, environmental randomness, and choice-sampling uniforms; realized choices and rewards can differ. Training, validation, and test participants are disjoint. Main results score test trials 151--200 or maps 7--8. Additional spatial prediction scores all eight test maps (Appendix~\ref{app:spatialA}).

\paragraph{Predictors and references.}
GRUs \citep{cho2014learning} and causal Transformers \citep{vaswani2017attention} are trained from scratch with three seeds each. LLaMA starts from the pretrained Llama-3.1-70B base model \citep{grattafiori2024llama}, not a Centaur-fitted checkpoint, and uses one fine-tuning seed. Its training follows the Centaur-style completion-masked pipeline, with 4-bit base weights and rank-8 adapters \citep{hu2021lora,dettmers2023qlora}. Every family is fitted offline on pooled participant sequences, with loss only at choice targets, causal access to preceding history, and no subject embeddings. Full-input and choice-only conditions use separate networks or adapters. No model is told the generating weight or participant parameters. Early stopping selects validation checkpoints. Tokenization, context units, and optimization details appear in Appendix~\ref{app:training}.

The \textbf{oracle} uses the generating policy and true participant parameters. A \textbf{fitted generator} estimates unknown parameters within the correct family from each test participant's first 150 restless trials or first six spatial maps, then predicts the remaining segment. It knows the policy family and fixed noise settings, but not the true individual parameters or weight. A simpler \textbf{RW + softmax} baseline fits one learning rate and inverse temperature to pooled training behavior at each weight; it has no choice kernel, spatial generalization, or fitted lapse. Its fitting regime matches pooled neural training, whereas the fitted generator is an individually calibrated reference. None of these models is refitted after reward replacement.

\paragraph{Comparability and uncertainty.}
All models receive the same actual choices, targets, and donor mappings. Spatial donor replacement preserves the initial cue; choice-only training omits all rewards, including that cue reward. GRU/Transformer curves average seed metrics, not ensemble predictions. Bands are 95\% participant-bootstrap intervals conditional on fits and donor maps, not training-seed uncertainty. Test participants do not select checkpoints or pooled parameters. Checkpoints, scoring windows, and primary operations were frozen before test evaluation; no test-based tuning or retraining was performed. Exploratory temporal-scope, deletion, and placeholder controls remain validation analyses (Appendix~\ref{app:operations}).

\section{Results}
\subsection{Restless choice histories retain nontrivial predictive information}
\label{sec:A}
\begin{figure}[H]
\centering
\includegraphics[width=\linewidth]{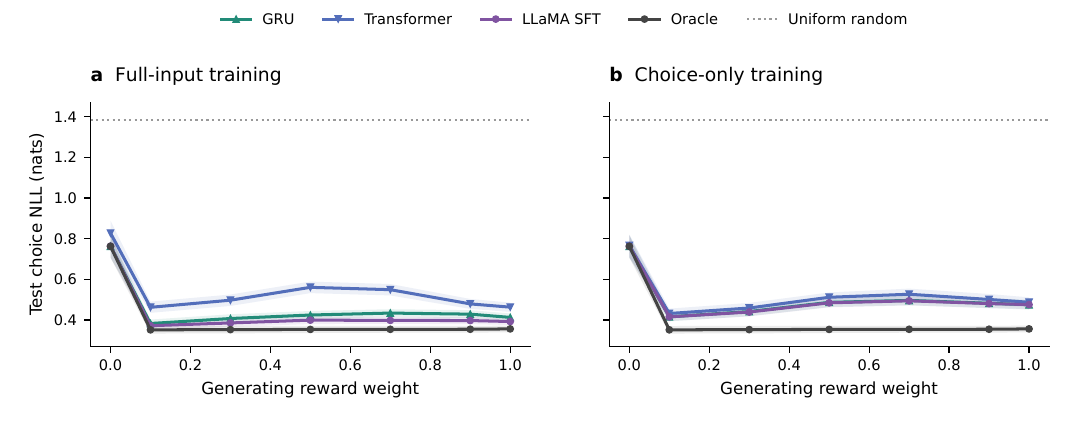}
\caption{\textbf{Choice-only prediction retains information in the restless task.} Restless-task NLL after full-input (left) or choice-only (right) training. The oracle observes rewards in \emph{both} panels; it is not a choice-only oracle. Dotted line: uniform random prediction. Curves use the same 250 test participants per weight and trials 151--200. Bands are participant-bootstrap intervals conditional on three GRU/Transformer fits and one LLaMA fit. Four train-fitted behavioral comparisons appear in Figure~\ref{fig:baselines}.}
\label{fig:prediction}
\end{figure}

Figure~\ref{fig:prediction} compares full-input and choice-only training in the restless task. All three neural families predict substantially better than uniform random choice in both settings. This remains true at $w=1$, where the generator has no explicit choice-kernel contribution. For example, LLaMA's mean NLL is $0.392$ with rewards and $0.475$ without them, compared with $\log 4=1.386$ for uniform prediction. Reward access helps, but reward omission does not eliminate predictability.

Beating random prediction alone would be insufficient: repeating the last choice can already perform well in reward-learning data. We therefore compare choice-only networks with training-set arm frequencies, a fitted stay probability, a first-order transition matrix, and a training-selected lag predictor. Baseline parameters and lag selection use training behavior only and remain frozen at test evaluation. All three neural families have lower mean NLL than each of these four baselines at every tested weight (Figure~\ref{fig:baselines}). This is a descriptive comparison across the tested families, not a claim that every possible behavioral heuristic has been excluded.

The defensible conclusion is that these networks extract predictive information beyond the specified simple summaries despite missing an input used by the generator. Whether that information amounts to implicit reward inference is not identified. Nor should the weight curve be read as a monotonic increase in predictability with reward contribution. The $w=0$ endpoint has a distinct tie structure, especially for negative kernels, which makes it intrinsically noisier than nearby positive weights (Appendix~\ref{app:zero}). Full-input prediction and choice-only prediction are therefore useful achievements, but neither is itself a mechanism-recovery test.

\subsection{Prediction and response-fidelity rankings diverge in the restless task}
\label{sec:B}
\begin{figure}[H]
\centering
\includegraphics[width=\linewidth]{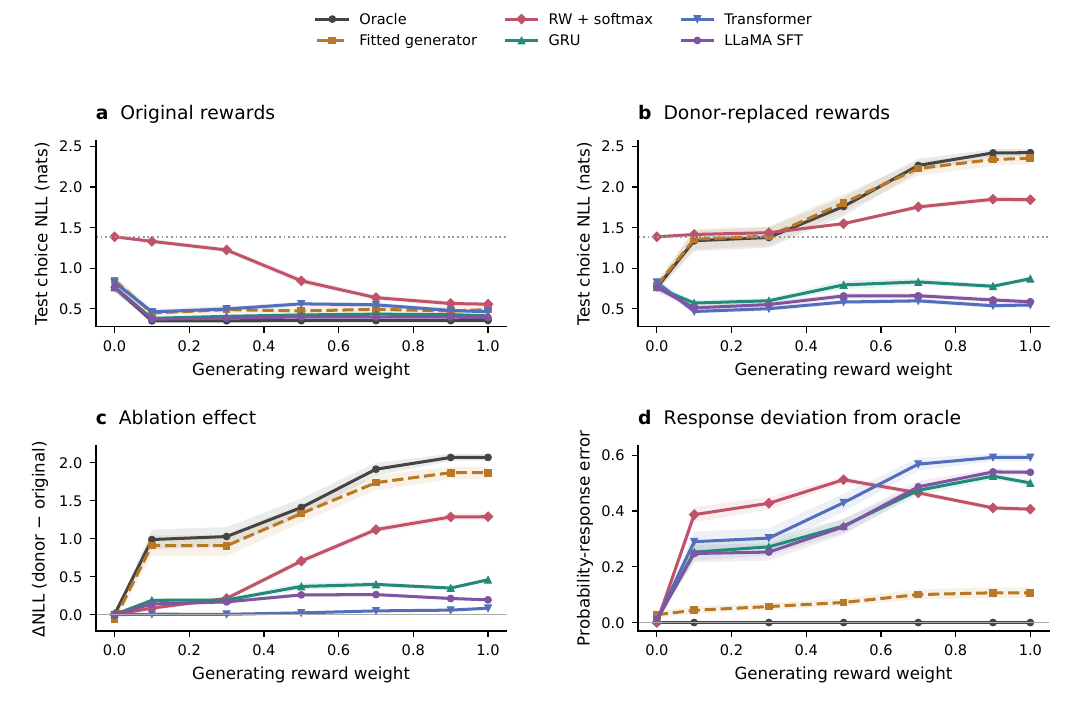}
\caption{\textbf{Better restless-task prediction need not imply a more faithful ablation response.} Original and donor-replaced NLL (a,b), their difference (c), and probability-response error relative to the oracle (d; Eq.~\ref{eq:error}). Actual choices are retained; 20 donor derangements replace preceding chosen rewards. High-weight neural predictors have better intact NLL than pooled RW but larger response error. Original/donor panels share limits. Bands are participant-bootstrap intervals; seed and calibration qualifications are as in Section~\ref{sec:methods}.}
\label{fig:restless}
\end{figure}

Figure~\ref{fig:restless} puts the intact and donor-replaced curves on the same scale before showing their differences and response errors. Reward replacement generally becomes more consequential as reward weight increases, particularly for the oracle. Yet the effect measured in the learned predictors can be far smaller than the effect in the process generating their data.

At $w=1$, the oracle's donor effect is $2.067$ nats. The corresponding effects are $0.458$ for GRU, $0.084$ for Transformer, and $0.194$ for LLaMA. These are not zero effects. Indeed, all three predictors benefit from reward information in training. The point is the size of their response relative to the matched reference. For LLaMA, a modest original NLL of $0.392$ coexists with a donor NLL of $0.586$, while the oracle's original NLL is about $0.356$ and its donor NLL is about $2.423$. Low reward sensitivity in the fitted model therefore does not entail weak reward dependence in this known generator.

The comparison with RW makes the distinction sharper. At $w=1$, RW predicts worse than each neural family: its NLL is $0.556$, versus $0.412$ for GRU, $0.463$ for Transformer, and $0.392$ for LLaMA. Nevertheless, its probability-response error is smaller: $0.407$, versus $0.501$, $0.592$, and $0.539$, respectively. The simpler predictor is not more accurate, but under this operation its probability changes are closer to the oracle's. For LLaMA versus RW, paired participant-bootstrap intervals support both directions at $w=.7,.9,1$: lower intact NLL and larger response error. These are unadjusted comparisons conditional on one LLaMA fit, not a general law about the model families.

The fitted-generator curve supplies a second reference. Its parameters are estimated from behavior, yet its response can remain close to the true-parameter oracle. That comparison distinguishes a known-family estimation problem from a flexible predictor's response without treating the estimated model as ground truth. Conversely, the pooled RW model is not entitled to preference merely because it is symbolic: at lower weights its approximation can be poor. The relevant finding is the dissociation between two rankings, not the universal superiority of one model class.

\subsection{The model ordering differs in the spatial task}
\label{sec:spatial}
\begin{figure}[H]
\centering
\includegraphics[width=\linewidth]{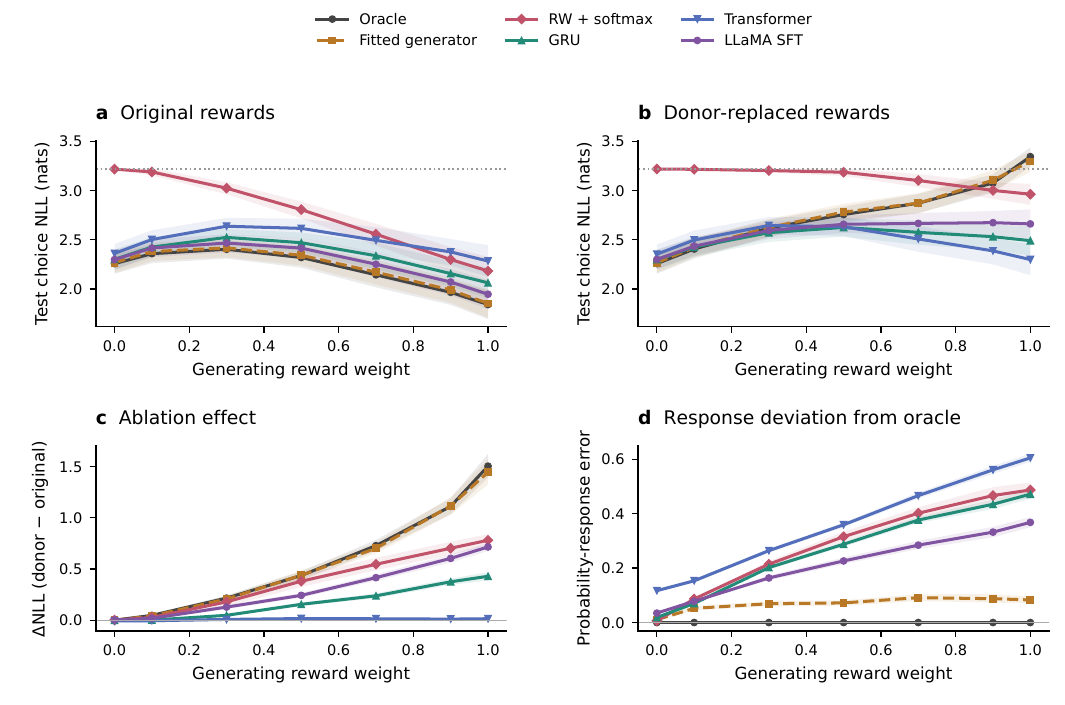}
\caption{\textbf{The spatial task shows a different predictor ordering.} Spatial-task test evaluation on maps 7--8: 50 participants per weight, 40 scored choices each, and 20 matched donor derangements. The initial location/reward cue remains intact. Panel definitions match Figure~\ref{fig:restless}; dotted lines mark $\log25$. Unlike in the high-weight restless task, LLaMA outperforms pooled RW in both original prediction and response fidelity at all tested nonzero weights. Only fixed-shape, causally audited LLaMA evaluations are included.}
\label{fig:spatial}
\end{figure}

The spatial task changes the reward structure, action space, and nonreward mechanism. Figure~\ref{fig:spatial} repeats the same four-panel comparison using the final two maps of each test participant. At $w=1$, the oracle's original NLL is $1.841$ and its donor effect is $1.504$ nats. The fitted generator is close on both quantities: $1.852$ and $1.444$. GRU, Transformer, and LLaMA again show smaller donor effects, respectively $0.429$, $0.015$, and $0.715$ nats. Thus the difference between strong prediction and oracle-sized sensitivity is not confined to the restless generator.

However, the neural-versus-symbolic ordering changes. At $w=1$, LLaMA predicts better than RW ($1.946$ versus $2.182$ NLL) \emph{and} has lower response error ($0.368$ versus $0.487$). The LLaMA advantage over RW in response error holds at every tested nonzero weight, with paired participant-bootstrap intervals excluding zero. Transformer remains much less sensitive than the oracle, whereas LLaMA retains a larger fraction of the operation-specific response. We therefore cannot describe the results as a necessary tradeoff between predictive flexibility and response fidelity. Both are empirical properties of a fitted pipeline in a particular task.

This is a replication of the \emph{evaluation framework} across a new mechanism combination, not zero-shot transfer: each model is trained separately on the spatial data. The task also changes the meaning of its mixture weight. The restless policy mixes utilities before an $\epsilon$-greedy decision, whereas the spatial policy mixes two probability distributions. Neither absolute NLL nor weight should be interpreted as a common mechanistic unit across the two tasks.

\section{Discussion}
\label{sec:discussion}
\paragraph{What does a small ablation effect establish?}
It establishes that the fitted predictor's measured performance is relatively insensitive to a specified input operation under a specified evaluation protocol. Our restless results show why an additional step is needed before interpreting that observation as a property of the generating behavior. Choices remain informative without rewards, and a model can exploit this information while expressing a much weaker response than the generator. The statement ``the model changes little, therefore the participants relied little on rewards'' is consequently not licensed by model ablation alone. This is a counterexample to an inference, not evidence that any particular human sample necessarily used a particular reward-learning mechanism.

\paragraph{A constructive calibration procedure.}
For reward-history ablations of sequential predictors, the practical alternative is to make the explanatory target explicit and test it independently of fit. First distinguish retraining without an input from perturbing an already trained predictor. Then report intact and perturbed performance on common targets, rather than only a relative percentage or an isolated loss difference. Finally, where a controlled simulator is available, compare probability responses under identical histories and operations, with a fitted same-family reference and a pooled simple baseline. Such calibration can reveal when predictive and response-fidelity rankings agree, reverse, or remain uncertain. When the human generator is unknown, a synthetic calibration cannot certify a human mechanism, but it can expose failures of the proposed inference before it is applied to human data. Other task classes require their own diagnostic input operations and references.

\paragraph{What the oracle does and does not resolve.}
The oracle defines a reproducible response for the chosen assay. It does not turn donor replacement into the causal effect of changing rewards in a live agent: the original choices remain fixed, including choices that would have changed under an altered environment. Donor histories also disrupt the joint structure of actions and rewards. The reference therefore calibrates the model against an explicitly defined conditional replay, not against a distribution-free measure of reward importance. Agreement is evidence about an observable response; disagreement cannot, by itself, locate the responsible latent computation. We intentionally avoid equating a recovered value trajectory, a scalar effect, and a recovered mechanism.

\paragraph{Structural alignment.}
RW shares a value-update motif with the restless generator, which may partly explain its high-weight response fidelity. Neural-generator controls change model rankings, but they also change behavior distributions and response scales; trained torsos inherit symbolic-data experience, and random teachers respond weakly to rewards (Appendix~\ref{app:neuralgenerator}). Thus they do not isolate generator-family bias or establish that simple cognitive models generally explain behavior better than neural models.

\paragraph{Limitations and future directions.}
More training seeds, especially for LLaMA SFT, and larger spatial samples would strengthen robustness assessment. Model differences cannot be ascribed to architecture alone: pretraining, tokenization, precision, optimization, and individual calibration access also vary. Future work should match these factors and test ablations beyond these sequential settings, including independent-trial decisions. Response fidelity is a validation criterion here, not a demonstrated training method.

\section{Conclusion}
Better prediction and more faithful input-ablation responses are distinct achievements in the sequential tasks studied here. Neural models predict from incomplete histories, yet their reward-perturbation responses can diverge from a known generator. A simpler model can predict worse but respond more faithfully, with rankings depending on the task and fitted pipeline. Interpreting these models as accounts of the generating process therefore requires separately validated responses, not accuracy alone.

\begin{ack}
We thank Robert C. Wilson for helpful discussions. This research was supported in part through research cyberinfrastructure resources and services provided by the Partnership for an Advanced Computing Environment (PACE) at the Georgia Institute of Technology, Atlanta, Georgia, USA.
\end{ack}

\section*{Code and data availability}
Code and frozen figure inputs are hosted at \url{https://github.com/xhb120633/transformer_bias_learning}. A matching code ZIP, synthetic data, test prediction exports, and trained model checkpoints are hosted at \url{https://huggingface.co/xhb120633/behavioral-model-ablations}. The LLaMA base model is not redistributed.

\clearpage
\appendix
\setcounter{figure}{0}
\renewcommand{\thefigure}{S\arabic{figure}}
\renewcommand{\theHfigure}{S\arabic{figure}}
\section{Generator specifications}
\label{app:generators}
\subsection{Restless reward learning with a signed one-step kernel}
For each arm $a$, the latent mean follows a clipped Gaussian random walk with mean reversion:
\begin{equation}
\mu_{t+1,a}=\operatorname{clip}_{[0,100]}\left[50+.9836(\mu_{t,a}-50)+\eta_{t,a}\right],\qquad
\eta_{t,a}\sim\mathcal N(0,2.8^2).
\end{equation}
Initial means are a participant-specific permutation of $(35,45,55,65)$. Potential rewards are sampled before choice, rounded to integers, and clipped to $[0,100]$ after adding Gaussian noise with standard deviation $4$. Only the chosen reward is revealed. Independent participant streams prevent recovery of a single shared deterministic schedule from the training set.

Initialize $Q_{i1,a}=50$. After observing $r_{it}$ at chosen arm $a_{it}$, update that arm by
\begin{equation}
Q_{i,t+1,a_{it}}=Q_{it,a_{it}}+\alpha_i(r_{it}-Q_{it,a_{it}}),
\end{equation}
leaving other arms unchanged. Learning rates follow $\alpha_i\sim\operatorname{Beta}(6,18)$. The kernel is $K_{it,a}=\mathbf{1}[a=a_{i,t-1}]$ for $t>1$ and zero initially. Let $s_i\in\{-1,+1\}$, sampled with equal probability, and $c_Q=14.819264931825328$. The policy utility and probabilities are
\begin{align}
u_{it,a}&=b_i\left[w\frac{Q_{it,a}-\bar Q_{it}}{c_Q}+(1-w)s_iK_{it,a}\right],\\
o_i(a\mid h_{it})&=.9\frac{\mathbf{1}[a\in M_{it}]}{|M_{it}|}+.1/4,
\qquad M_{it}=\arg\max_a u_{it,a}.
\end{align}
The positive scale $b_i$ is sampled from a log-normal distribution with median $2.25$ and log standard deviation $.15$. It cancels in this $\epsilon$-greedy policy and is not an identifiable behavioral parameter here. The reward scale was frozen using independent random-policy calibration trajectories, not fit on evaluation targets. The kernel sign penalizes or favors repetition; a negative sign is not a negative learning rate and does not, by itself, implement uncertainty-directed exploration.

\subsection{Why the zero-weight endpoint is distinctive}
\label{app:zero}
For a positive kernel at $w=0$, the last choice uniquely maximizes utility. For a negative kernel, the three other choices tie. The policy is therefore $0.025$ on the last choice and $0.325$ on each alternative. At a positive reward weight, learned values can break these ties even if the kernel still suppresses staying. A large change between $w=0$ and $w=.1$ can consequently reflect a change in the number of favored options, not an abrupt discovery of reward information by a predictor. At $w=1$, exact value ties are uncommon after learning; a unique winning arm receives probability $.925$ and the others $.025$. These policy differences affect attainable NLL and prohibit a simple identification of weight with entropy, prediction difficulty, or ablation magnitude.

\subsection{Spatial generalization and local search}
Each map samples latent payoffs on the 25 grid positions from a Gaussian process with mean $50$ and covariance
\begin{equation}
\operatorname{Cov}(f(x),f(x'))=15^2\exp\left(-\frac{\|x-x'\|_2^2}{2(1.2)^2}\right).
\end{equation}
Observed rewards add independent Gaussian noise with standard deviation $2$, without clipping or rounding. The learner uses normalized rewards $(r-50)/15$, zero prior mean, unit signal variance, observation variance $(2/15)^2$, and an RBF kernel with participant-specific length $\ell_i$. It updates the Gaussian posterior from the initial cue and each chosen reward. Let $m_{it}(a)$ denote the posterior mean, and $d_1(a,a_{i,t-1})$ Manhattan distance from the last visited position (the cue position initially). Then
\begin{equation}
o_i(a\mid h_{it})=.9\left[w\softmax\left(\frac{m_{it}}{\tau_{R,i}}\right)_a
+(1-w)\softmax\left(-\frac{d_1(\cdot,a_{i,t-1})}{\tau_{L,i}}\right)_a\right]+.1/25.
\end{equation}
Independently for each participant, $\ell_i\sim U(.8,1.6)$, $\tau_{R,i}\sim U(.12,.35)$, and $\tau_{L,i}\sim U(.5,1.1)$. Each participant keeps these parameters across eight independent maps. The reward-based component uses posterior means only, with no UCB uncertainty bonus. The task is inspired by spatial exploration \citep{wu2018generalization}, but the mixture and reduced grid are designed for controlled synthetic calibration.

\section{Generator checks and recovery limits}
\label{app:generatorchecks}
\begin{figure}[t]
\centering
\includegraphics[width=\linewidth]{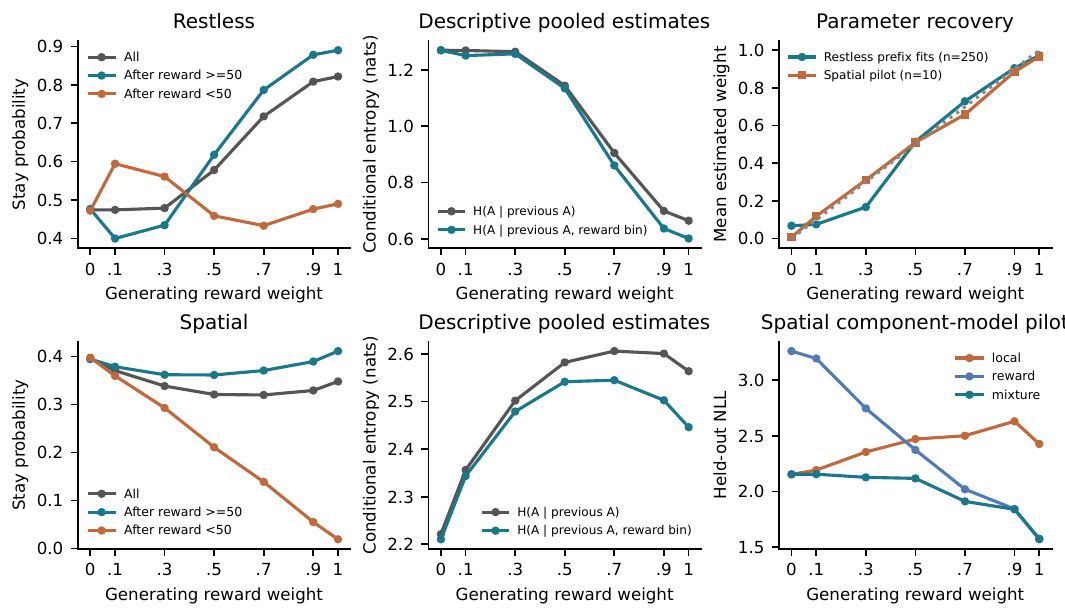}
\caption{\textbf{Behavioral checks and bounded recovery evidence.} Left: repetition probabilities in existing training records, including conditioning on preceding reward. Middle: pooled plug-in conditional entropies, not held-out information estimates. Top right: mean fitted reward weights from 250 restless validation-prefix fits and a separate spatial pilot with 10 training participants per weight. Dashed line denotes equality; means can conceal substantial individual error and nonidentifiability (Figure~\ref{fig:recovery}). Bottom right: spatial pilot NLL on two maps withheld from individual fitting, comparing local-only, reward-only, and mixture candidates. Curves are descriptive; no training-seed or inferential bands are implied.}
\label{fig:generatorchecks}
\end{figure}

\paragraph{Behavioral manipulation checks.}
Figure~\ref{fig:generatorchecks} summarizes the existing training data: 1,000 restless and 200 spatial participants per weight. We measure repetition, repetition following high versus low rewards, and pooled conditional entropy. Because rewards are continuous or multivalued rather than binary wins/losses, we define a descriptive high-reward indicator $B_{t-1}=\mathbf{1}[r_{t-1}\geq50]$, using the environments' common prior mean rather than choosing a threshold from results. High-reward stay and low-reward shift are the corresponding win-stay/lose-shift analogues; low-reward shift is one minus the plotted low-reward stay rate. Transitions across map boundaries and the first choice of each session/map are excluded.

We compute empirical frequency estimates of $H(A_t\mid A_{t-1})$ and $H(A_t\mid A_{t-1},B_{t-1})$, in nats. Their difference is a descriptive conditional mutual information, not a causal effect or the information in the full reward history. These pooled plug-in estimates have finite-sample bias and mix individual policies; they are not held-out prediction baselines or additional hypothesis tests. In particular, conditioning is on the \emph{preceding} reward, never the outcome of the choice being predicted.

In the restless data, mean obtained reward rises from $49.915$ at $w=0$ to $60.890$ at $w=1$, while the two conditional entropies change from $1.269/1.269$ to $.665/.602$. The spatial endpoints are $50.384$ versus $63.890$ for reward and $2.221/2.210$ versus $2.564/2.446$ for the entropies. Thus increased reward contribution does not impose the same entropy trajectory across tasks. Nor must pooled high-reward stay exceed low-reward stay: at restless $w=.1$, the negative-kernel group earns higher rewards while almost always switching, whereas the positive-kernel group almost always stays. Pooling them reverses the apparent reward--stay association. Within those groups, stay rates are $.025$ and $.925$, respectively. These summaries characterize the manipulation but cannot independently identify its mechanisms.

\begin{figure}[p]
\centering
\includegraphics[width=\linewidth]{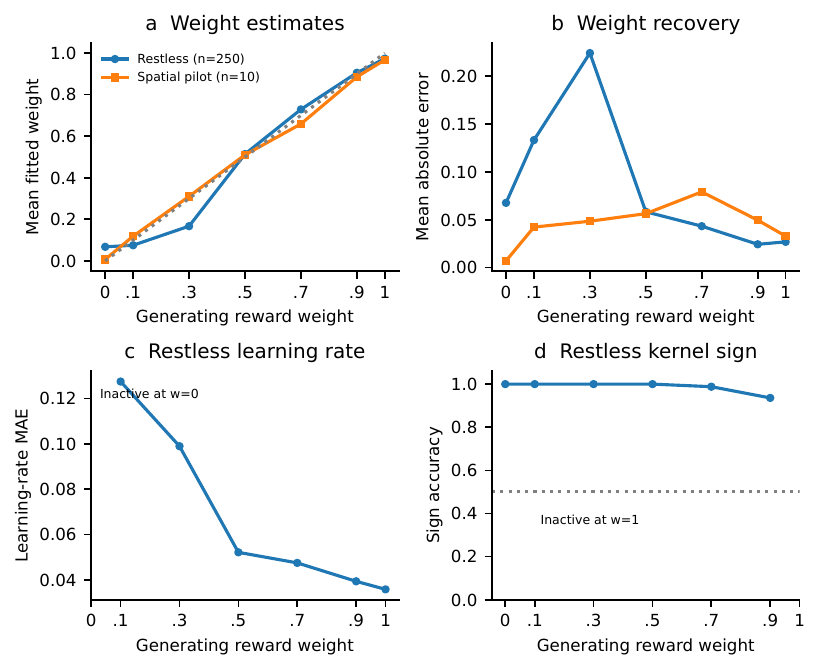}
\caption{\textbf{Parameter recovery is not uniform.} (a) Mean estimated reward weight and identity line; (b) weight mean absolute error (MAE); (c) restless learning-rate MAE; (d) restless kernel-sign accuracy, with chance marked at $.5$. Restless metrics use validation-participant prefix fits ($n=250$ per weight); spatial estimates use a training-participant pilot ($n=10$ per weight). Inactive parameters at $w=0$ or $w=1$ are omitted, not scored as zero error. Curves are descriptive without uncertainty bands; mean agreement in (a) need not imply accurate individual recovery in (b).}
\label{fig:recovery}
\end{figure}

\paragraph{Restless parameter recovery from existing fits.}
This exploratory recovery analysis retains the earlier validation-prefix fits, not the independent-test fits in the main figures. Each validation participant's first 150 trials estimate $\alpha,w,s$ without access to ground-truth parameters. The grid has 82 learning-rate values, 101 weights, and two signs; noise and reward scaling are known. Figure~\ref{fig:recovery} evaluates the first selected optimum, retaining the original deterministic tie-breaking. At low weights, weight and learning-rate recovery can be poor despite known-family fitting. Between $85.2\%$ and $97.2\%$ of participants have multiple exactly tied grid optima across conditions, consistent with the piecewise-constant likelihood of this $\epsilon$-greedy policy. Mean weight spans among candidates within one total prefix NLL unit of the optimum are $.252,.437,.340,.083,.082,.075,.069$ in weight order. These spans describe optimization ambiguity; they are not confidence intervals. At $w=0$ the learning rate is inactive, and at $w=1$ the kernel sign is inactive, so their recovery is not scored. Correct-family response fidelity must therefore be assessed directly rather than inferred from an assumed unique parameter estimate.

\paragraph{Spatial component-model recovery pilot.}
The existing pilot uses the first 10 training participants per weight, fitting each participant's first six maps and evaluating the final two. It compares local-only ($w=0$), reward-only ($w=1$), and mixture candidates with one, two, and four free parameters, respectively. Three-start bounded L-BFGS-B fitting uses known observation noise and lapse; all selected fits report optimizer success. BIC selects local-only for all 10 participants at $w=0$, mixture for all 10 at each of $w=.3,.5,.7$, and reward-only for all 10 at $w=1$. Near endpoints it often selects the simpler candidate: six local-only at $w=.1$ and eight reward-only at $w=.9$. Held-out NLL likewise does not always select the generating component model (Figure~\ref{fig:generatorchecks}). This is a limited nested-component selection check, not a comprehensive confusion matrix across unrelated generating families. It uses neither validation nor test participants and does not establish recoverability of every nuisance parameter. Together with the restless ambiguities, it supports treating known weights as design controls, not as quantities guaranteed to be uniquely recoverable from short choice sequences.

\section{Training and cognitive fitting}
\label{app:training}
\subsection{Neural representations and optimization}
In the restless task, scratch-trained networks receive a beginning-of-sequence token followed by separate choice and reward-value tokens. A choice/reward pair is never encoded as one compound token. Integer rewards have their own vocabulary entries. Full records contain 401 tokens for 200 trials; choice-only records contain the beginning token and 200 choices. Training predicts choice tokens causally from the accumulated prefix, with no loss on reward targets. Full sequences are processed jointly, not optimized one trial at a time. The GRU is unidirectional and the Transformer uses a causal attention mask.

In the spatial task, scratch-trained models receive a symbolic grid description, cue position, and the 20-trial map record. Rewards are encoded through numeric characters using round-trip float strings rather than reward bins. Every independent map is a separate sequence. Choice-only records preserve the grid and cue position but omit both cue and subsequent reward observations. Models do not receive previous maps as context. Thus pooled neural fitting can learn history-dependent computations within a map, but it has no cross-map participant identifier or embedding.

\begin{table}[h]
\caption{Scratch-trained neural configurations. Seeds are 11, 22, and 33 in each weight/input condition. Batch sizes are effective sequence batch sizes.}
\label{tab:hyper}
\centering\small
\begin{tabular}{lll}
\toprule
Setting & GRU & Transformer\\
\midrule
Embedding/model width & 64 / 256 hidden & 256\\
Layers & 2 & 6\\
Attention heads / feed-forward width & Not applicable & 8 / 1,024\\
Dropout & .1 & .1\\
Optimizer & AdamW & AdamW\\
Learning rate / weight decay & $.001$ / $.01$ & $.0003$ / $.01$\\
Effective batch size & 64 & 128\\
Maximum epochs / patience & 200 / 20 & 60 / 8\\
Stopping minimum improvement & $.0001$ & $.0002$\\
Warm-up steps & 32 & 150\\
Restless / spatial precision & FP32 / FP32 & BF16 / FP32\\
\bottomrule
\end{tabular}
\end{table}

LLaMA uses \texttt{unsloth/Meta-Llama-3.1-70B-bnb-4bit}, with rank-stabilized LoRA enabled, rank 8, alpha 8, and zero adapter dropout. Adapters target attention query, key, value, and output projections, plus gate, up, and down projections in the feed-forward blocks. The optimization uses 8-bit AdamW, learning rate $5\times10^{-5}$, weight decay $.01$, cosine scheduling, 100 warm-up steps, batch size one, and 32-step gradient accumulation. Training runs for at most 20 epochs with patience three, evaluation each epoch, and restoration of the best validation checkpoint. The seed is 100. Maximum context capacity is 32,768 tokenizer tokens, not trials. Natural-language instructions describe the task without revealing the weight or parameters. Choice delimiters mark supervised completions; reward and instruction text supplies context rather than training targets. We preserve this pipeline across full-input and choice-only training. It is a task-specific SFT experiment using the base model, not an evaluation of the published Centaur adapter.

Reported NLL uses choice probabilities conditional on the four or 25 valid action tokens. We retain raw full-vocabulary likelihood and valid-choice mass for auditing; these are not silently interchanged. The main figures use uncalibrated conditional probabilities, not post-hoc temperature-adjusted scores. Since LLaMA uses natural-language tokenization and large-scale pretraining whereas the small networks do not, the study compares modeling pipelines rather than a controlled scaling experiment.

\subsection{Correct-family fitting}
For each restless test participant, the fitted-generator reference minimizes NLL on trials 1--150 over the learning rate, reward weight, and kernel sign. The exact $\epsilon$-greedy likelihood is piecewise constant in parts of parameter space, so fitting uses a deterministic grid rather than gradients: 81 evenly spaced learning rates from $.01$ to $.99$ plus $.25$, 101 weights from zero to one, and both signs. The first minimum in the fixed grid order defines the plotted fit. A grid optimum is an approximation, not proof of a unique parameter estimate. The noise level and value scale are fixed to their generating values. Only the first segment enters estimation; fitted parameters remain frozen for subsequent intact and donor replay. Replacing rewards in the calibration segment at evaluation does not erase the information already retained in these fitted parameters.

For spatial participants, we fit the mixture family on maps 1--6 and evaluate maps 7--8. Unknown parameters are GP length, reward temperature, local temperature, and mixture weight. Their bounds are $[.2,4]$, $[.03,2]$, $[.15,3]$, and $[0,1]$, respectively. Positive parameters are optimized in log space. Three L-BFGS-B starts use lengths/temperatures $(1,.25,.8)$ with weights $.15,.5,.85$, a maximum of 150 iterations, and function tolerance $10^{-10}$. Noise and lapse remain fixed. The selected fits pass optimizer-success checks. These individual reference fits are not comparable in data access to pooled students without that qualification, and are not described as choice-only oracles.

The main test reference uses the fine grid above. Earlier validation reference curves used 41 evenly spaced learning rates plus $.25$; the supplementary parameter-recovery analysis used the fine grid. Changes in fitted-generator curves between splits therefore also include grid resolution, not only new participants.

\subsection{Pooled RW baseline}
RW uses the same chosen-arm delta rule, initial normalized value zero, and reward transform $(r-50)/100$, with $p(a)\propto\exp(\beta Q_a)$. In the spatial task, it updates on the initial cue and resets at each map. It has no spatial kernel and cannot generalize an observation to nearby arms. One $(\alpha,\beta)$ pair is fit per weight and task using all training participants. Nine L-BFGS-B starts cross $\alpha\in\{.05,.3,.8\}$ and $\beta\in\{.5,5,50\}$, using analytic gradients, bounds $\alpha\in[0,1]$ and $\beta\in[10^{-4},10^3]$, and at most 300 iterations. The best converged training likelihood is selected, with checks against better unconverged starts. Parameters never adapt on validation or test targets, but the model's values update from each observed history, as do the neural predictors' conditional states. This distinction between fixed parameters and changing history-dependent states applies to all families.

\section{Behavioral controls for choice-only prediction}
\label{app:baselines}
\begin{figure}[t]
\centering
\includegraphics[width=\linewidth]{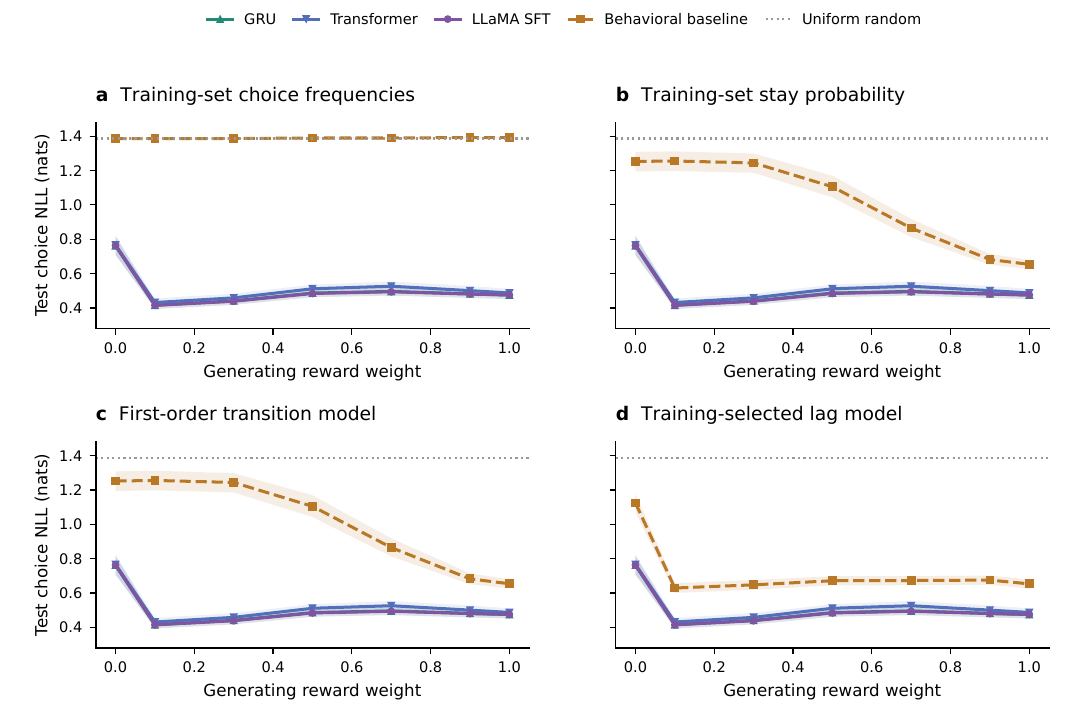}
\caption{\textbf{Choice-only prediction exceeds four simple training-fitted controls.} Each panel repeats the Figure~\ref{fig:prediction} choice-only neural curves and adds the named behavioral baseline. All baseline parameters, including lag, are selected on training data and frozen at test evaluation. Comparisons use the same 250 participants and trials 151--200. Uniform random prediction is dotted. Neural mean NLL is lower in every displayed model/weight comparison; this is not a multiplicity-adjusted significance claim.}
\label{fig:baselines}
\end{figure}
The static-frequency model predicts the pooled training proportion of each arm. The stay model assigns the pooled training repeat probability to the immediately preceding choice and distributes the remaining probability uniformly over the other three arms. The transition model estimates a $4\times4$ matrix with unit Laplace pseudocounts and conditions on the preceding actual choice. The lag model considers lags 1--20; for each lag it fits the probability of repeating that earlier choice, again allocating the remainder uniformly. Training NLL selects the lag, with a uniform prediction when the requested lag is unavailable. Candidate comparisons use the same training-trial range. The selected lag is two for weights through $.9$ and one at $w=1$. Neither validation nor test feedback changes baseline parameters or the selected lag. Using an earlier observed choice as an input is causal conditioning, not refitting.

\section{Additional spatial prediction results}
This is an additional training-ablation result, not a duplicate of the main figures: Figure~\ref{fig:prediction} compares full-input versus choice-only training in the restless task, whereas Figure~\ref{fig:spatial} evaluates fixed full-input models under reward replacement in the spatial task.
\label{app:spatialA}
Table~\ref{tab:spatialA} reports the completed LLaMA full-input/choice-only comparison on all eight spatial maps. This wider scoring window differs from the last-two-map ablation comparison, so its NLL must not be subtracted from the donor NLL in Figure~\ref{fig:spatial}. At positive weights, reward access consistently improves LLaMA's mean prediction; choice-only prediction remains far below the uniform baseline $\log25=3.219$. These results establish prediction under a reduced input set. The four restless behavioral controls were not validated as spatial baselines, so we do not claim that this table rules out the corresponding range of spatial heuristics.
\begin{table}[h]
\caption{Spatial LLaMA prediction on 50 test participants, all 160 choices each. One full-input and one choice-only adapter per weight. Values are NLL in nats.}
\label{tab:spatialA}
\centering
\begin{tabular}{rrrr}
\toprule
Reward weight & Full input & Choice only & Choice only $-$ full\\
\midrule
0.0 & 2.279 & 2.280 & 0.001\\
0.1 & 2.386 & 2.393 & 0.006\\
0.3 & 2.448 & 2.525 & 0.077\\
0.5 & 2.389 & 2.512 & 0.123\\
0.7 & 2.273 & 2.441 & 0.169\\
0.9 & 2.117 & 2.339 & 0.222\\
1.0 & 1.983 & 2.223 & 0.240\\
\bottomrule
\end{tabular}
\end{table}

\clearpage
\section{Sensitivity to generator construction}
\label{app:neuralgenerator}
\subsection{Trained GRU torso with reset output heads}
These exploratory controls retain validation evaluation. A GRU torso trained on reward-dominant restless data is frozen with output heads reset using seeds 101, 102, and 103. Two copies with distinct hidden states process each realized choice history, receiving actual rewards or constant 50. Their policies are mixed across seven weights, and teacher-sampled choices determine later histories. The exact teacher is the oracle.

\begin{figure}[H]
\centering
\includegraphics[width=\linewidth]{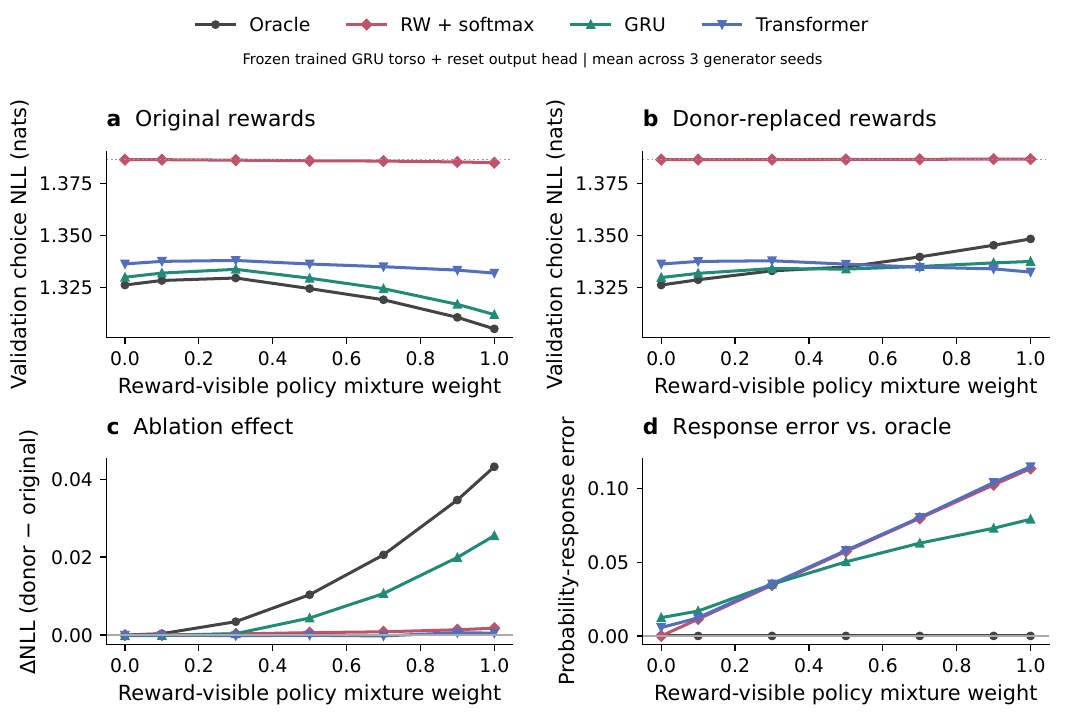}
\caption{\textbf{A neural-generator sensitivity control.} Validation evaluation with a frozen previously trained GRU torso and reset output heads; each curve averages three generator seeds. Four panels match the main response comparisons, but the horizontal axis is a reward-visible versus constant-reward \emph{policy-mixture} weight, not the original psychological utility weight. Students are freshly fitted; no student parameters are shared with the teacher. This experiment does not use LLaMA and is not a fully random-generator control.}
\label{fig:neuralteacher}
\end{figure}

For each of the 21 cells, we generate 500 training and 100 validation sessions of 200 trials. Fresh GRU and Transformer students, with full-input and choice-only variants, are trained with one student seed per cell; a pooled RW baseline is also fitted. The response comparison uses trials 151--200 and 20 matched donor permutations. Figure~\ref{fig:neuralteacher} averages over the three generator heads. It shows neither individual-seed traces nor uncertainty bands.

At mixture weight one, mean intact NLL is $1.305$ for the teacher, $1.385$ for RW, $1.312$ for GRU, and $1.332$ for Transformer. Mean donor effects are $.043$, $.002$, $.026$, and $.0005$, respectively; response errors are zero, $.114$, $.079$, and $.115$. GRU predicts better than RW in all 21 cells, but has lower response error in only 12. This does not establish an architecture-matching law: the generator inherited a trained torso, and its response magnitude and behavior distribution differ from the symbolic generator.

\clearpage
\subsection{Trained Transformer torso with reset output heads}
We construct the corresponding Transformer control from the reward-weight-one, seed-11 full-input checkpoint. Its original readout shares weights with the input token embedding. We therefore detach that readout and initialize a new bias-free linear output layer, using seeds 101, 102, and 103, before freezing the teacher. All remaining parameters, including the learned input embedding, are verified unchanged. The reset uses the linear layer's default initialization; no heads are selected or rescaled based on behavioral outcomes. The teacher processes the complete causal prefix and mixes actual-reward and constant-50 policies as above. Each of the 21 cells uses 500 training and 100 validation sessions, the same student families and early-stopping procedure, and 20 donor permutations scored on trials 151--200. Generation/replay and causal-prefix checks pass for all three teachers. No LLaMA students are used.

\begin{figure}[H]
\centering
\includegraphics[width=\linewidth]{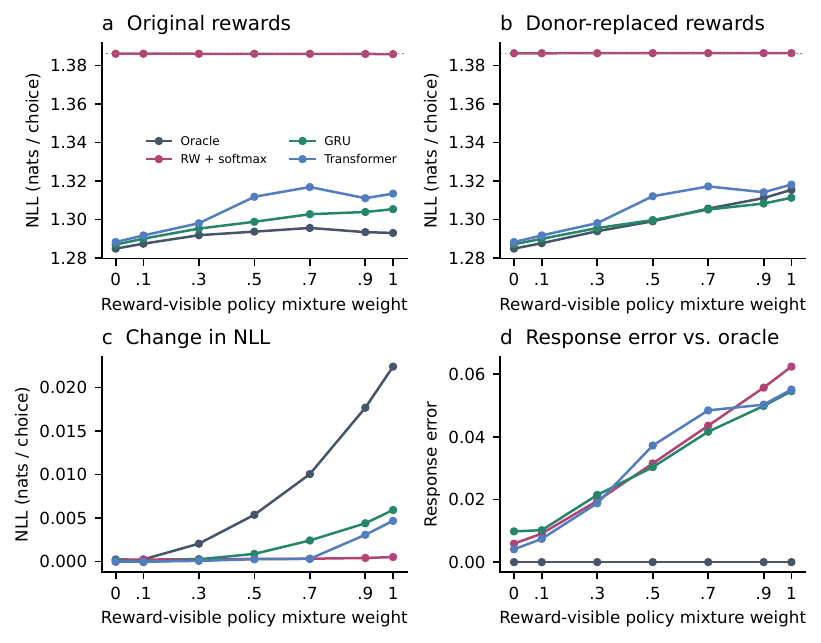}
\caption{\textbf{Trained Transformer torso with reset output heads.} Validation curves average three head seeds, without individual-seed traces or uncertainty bands. Panel definitions match the trained-GRU control. The new output layer is untied from the learned token embedding, leaving the torso unchanged. Original and donor NLL share limits; dotted lines mark uniform prediction. This is a trained-representation control, not a fully random Transformer.}
\label{fig:trainedtransformer}
\end{figure}
At mixture weight one, intact NLL is $1.2931$ for the teacher, $1.3859$ for RW, $1.3054$ for GRU, and $1.3135$ for Transformer. Their donor effects are $.0224$, $.0005$, $.0059$, and $.0047$, respectively. GRU and Transformer response errors ($.0546$ and $.0551$) are below RW's $.0624$ at this endpoint. Across the 21 cells, both students predict better than RW, but have lower response error in only 11 and 15 cells, respectively. These descriptive counts do not establish statistical significance or an architecture-matching advantage. The trained torso retains experience from symbolic data, and this construction changes both the behavioral distribution and oracle response scale.

\clearpage
\subsection{Fully randomly initialized GRU}
We repeat the procedure with a GRU whose \emph{entire} parameter set is randomly initialized and frozen, with no checkpoint loaded. It uses seeds 101, 102, and 103 and the same seven policy-mixture weights, sample sizes, student-fitting regime, scoring window, and 20 donor assignments specified above. The GRU maintains recurrent state and mixes its actual-reward and constant-50 policies. All 21 cells are retained, without seed selection, temperature tuning, or post-hoc rescaling. Generation/replay agreement, unchanged first-choice probabilities, future-reward prefix invariance, and reward invariance at zero weight pass for all three teachers. This control uses no LLaMA students.

\begin{figure}[H]
\centering
\includegraphics[width=\linewidth]{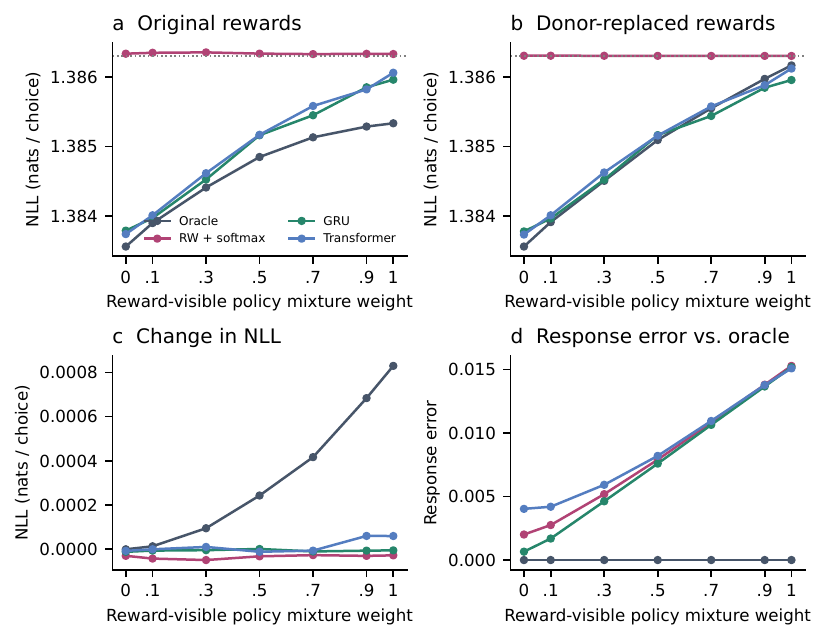}
\caption{\textbf{Fully random GRU generator.} All weights, not just the output head, are initialized independently of behavioral data. Validation curves average three generator seeds, with one student seed per cell. Panel definitions and scoring window match the trained-torso control. Dotted lines mark uniform NLL; original and donor panels share limits. The oracle's very small effect sets the scale of this control, not evidence of successful neural recovery.}
\label{fig:randomgru}
\end{figure}
At weight one, the random-GRU oracle has NLL $1.38533$, close to $\log4=1.38629$, and a donor effect of only $.00083$ nats. GRU, Transformer, and RW response errors are $.01519$, $.01510$, and $.01530$. Thus small absolute errors here occur in a weak-response regime. This control has limited power to discriminate response fidelity and cannot establish the absence of generator bias.

\clearpage
\subsection{Fully randomly initialized Transformer}
We apply the same fully random initialization and evaluation protocol to a Transformer, again using seeds 101, 102, and 103 across seven policy-mixture weights. Unlike the recurrent teacher, it processes the complete causal prefix. All 21 cells are retained without seed selection, temperature tuning, or post-hoc rescaling, and the same four causal and replay checks pass for all three teachers. Student families, sample sizes, scoring windows, and donor counts match the random-GRU control; no LLaMA students are used.
\begin{figure}[H]
\centering
\includegraphics[width=\linewidth]{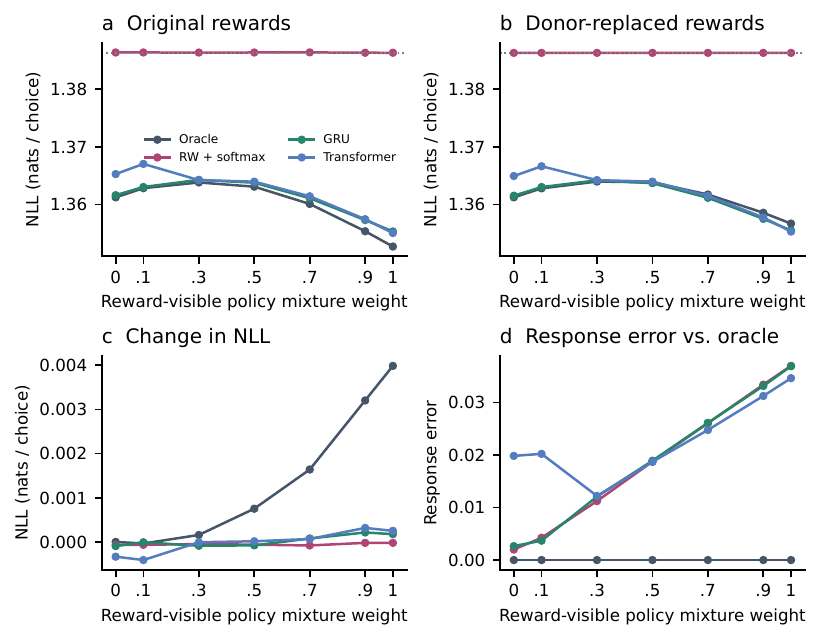}
\caption{\textbf{Fully random Transformer generator.} Validation protocol and panel definitions match Figure~\ref{fig:randomgru}; all 21 cells are included. Each student is trained on this teacher's own generated choices, not the symbolic generator's choices. No individual-seed traces or uncertainty bands are shown. Axis ranges differ from the random-GRU figure because the generators induce different prediction and response scales.}
\label{fig:randomtransformer}
\end{figure}
For the random Transformer at weight one, oracle NLL is $1.35270$ and its donor effect is $.00398$. GRU and Transformer students predict better than RW ($1.35532$ and $1.35504$ versus $1.38632$), but their donor effects remain small ($.00018$ and $.00025$). Response errors are $.03694$, $.03465$, and $.03699$ for GRU, Transformer, and RW. Changing the generator therefore changes the comparison's scale and ordering. These exploratory controls remove inherited trained weights, but do not hold the behavioral distribution or oracle response strength fixed; they do not isolate architecture alone.

\clearpage
\section{Sensitivity to the reward-perturbation operation}
\label{app:operations}
\paragraph{Matched seven-weight temporal-scope controls.}
The main operation permutes complete participant reward sequences while retaining their internal temporal order. It is not a shuffle of trial order within a participant. In an exploratory validation analysis at the same seven weights, using 20 donor maps and trials 151--200, we compare replacing (i) all preceding rewards, (ii) only the immediately preceding reward for each target, or (iii) preceding rewards from trial 151 onward. For (ii), each target starts from its unperturbed earlier history, so replacements do not accumulate. For (iii), the first 150 rewards remain intact. Parameters stay fixed in every case. The correct-family fit retains its first-150-trial estimates even in (i).

\begin{figure}[H]
\centering
\includegraphics[width=\linewidth]{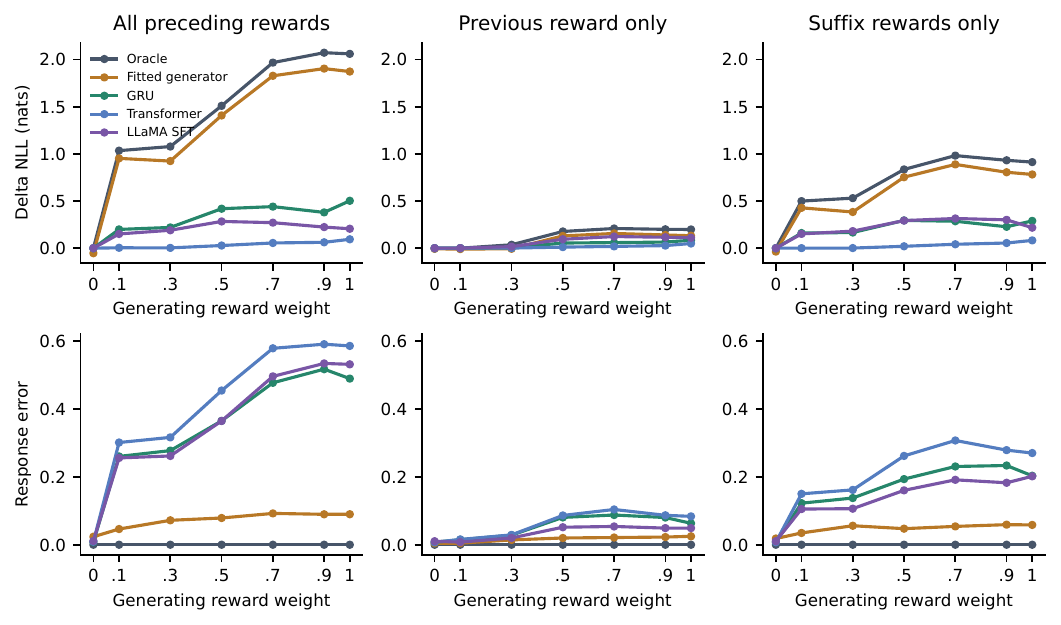}
\caption{\textbf{The scope of reward replacement changes the measured effect.} Columns vary which preceding rewards are replaced; rows report NLL change and response error against the oracle under that same operation. Curves average three GRU/Transformer seeds and use one LLaMA SFT seed; oracle and individual correct-family fits use the same participants and donors. Shared limits within each row aid comparison. These are descriptive means from the existing seven-weight validation analysis.}
\label{fig:scope}
\end{figure}
These operations target different conditional responses. A small single-reward effect cannot substitute for an all-history effect, and none is the closed-loop effect of changing rewards in an acting agent. Deletion and nonnumeric placeholders do not provide numerical rewards to a symbolic update rule without an additional missing-data convention, so they are not assigned the same oracle reference by default.

\begin{figure}[H]
\centering
\includegraphics[width=\linewidth]{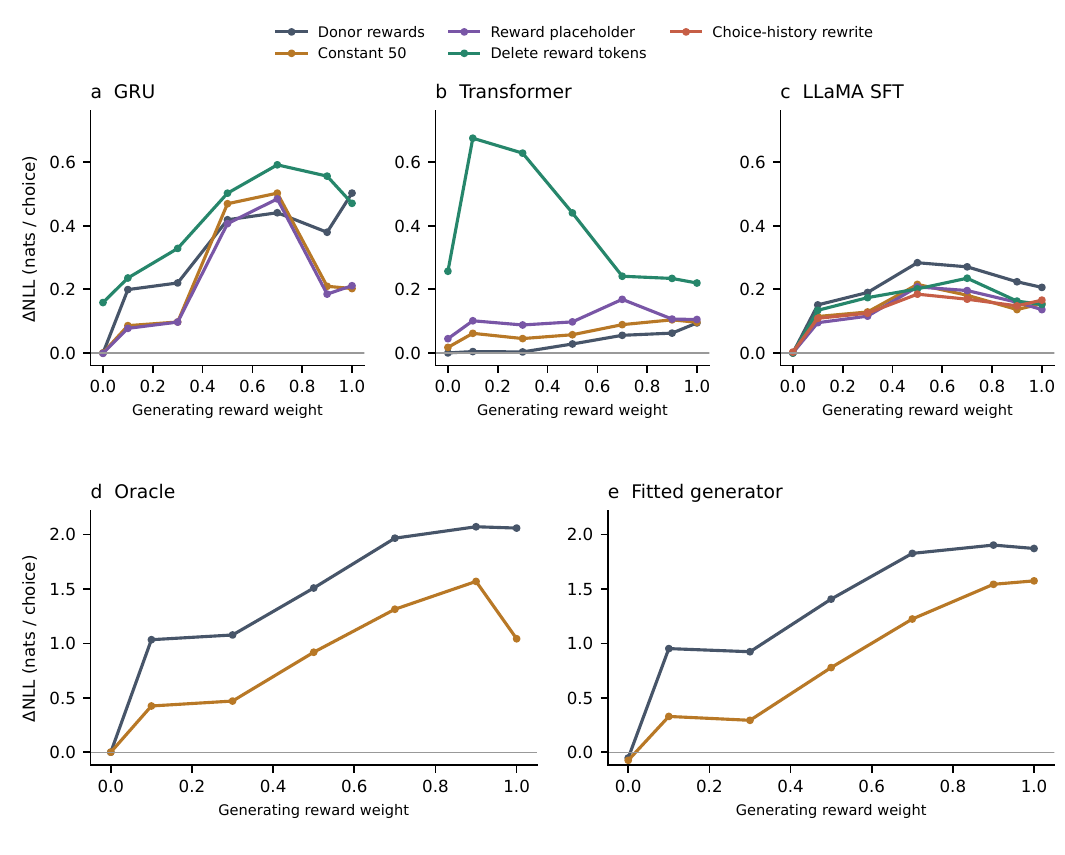}
\caption{\textbf{Restless-task reward-edit effects depend on the model and operation.} Seven-weight validation sweep of fixed full-input GRU (a), Transformer (b), and LLaMA SFT (c) models, followed by the true-parameter oracle (d) and prefix-fitted generator (e). Each point averages 250 participants on trials 151--200, over three GRU/Transformer seeds or one LLaMA SFT seed. The donor curves retain the 20 validation assignments, not the main test assignments. Constant 50 is replayed through both symbolic policies; deletion and placeholders have no symbolic response without a missing-input convention. The choice-history rewrite applies only to LLaMA. Vertical limits match within each row; rows have different scales. Means are shown without seed traces or uncertainty bands.}
\label{fig:operationsweep}
\end{figure}

\clearpage
\paragraph{Matched seven-weight reward-operation controls.}
Figure~\ref{fig:operationsweep} compares reward edits in the same seven-weight restless-task validation data: 250 participants per weight, scored on trials 151--200. We reuse full-input GRU and Transformer checkpoints (three seeds per family) and LLaMA SFT adapters (one seed), with parameters fixed throughout evaluation. The oracle and individually fitted generator provide numerical-reward references. Every panel reports four-choice-normalized NLL. This comparison replaces the earlier three-condition exploratory display.

We compare donor rewards, constant reward 50, a reward placeholder, and deletion of reward tokens in the neural models. GRU and Transformer placeholders zero the reward-token input embeddings; LLaMA placeholders use a reserved special token. These preserve token positions, although their representations differ. Deletion shifts remaining tokens and prediction positions. LLaMA also receives a choice-history rewrite, which changes the instruction and removes reward-bearing feedback. For the oracle and fitted generator, donor replacement and constant 50 are both valid numerical inputs to the original update rule; we replay both on actual choices with parameters fixed after fitting. Missing reward tokens and nonnumeric placeholders require a separately specified update convention and therefore have no symbolic curve here. Choice-only retraining remains a separate experiment.

At weight one, GRU's donor and deletion effects are $.502$ and $.470$ nats per choice; Transformer shows $.094$ and $.220$; LLaMA SFT shows $.206$ and $.152$. The oracle's donor effect is $2.060$, while its constant-50 effect is $1.043$; corresponding fitted-generator effects are $1.873$ and $1.575$. Even the known generator therefore gives different effect magnitudes under two valid reward edits. Transformer deletion also has an effect of $.257$ at weight zero, where the oracle's reward-policy contribution is zero. Neither the size nor the ordering of an edit effect can be read directly as the generating reward weight.

\section{Numerical and provenance checks}
\label{app:audit}
For the seven-weight LLaMA operation sweep, an initial deletion evaluation changed first-choice log probabilities by up to $.0952$ despite an unchanged causal prefix. Those exports are excluded from Figure~\ref{fig:operationsweep}. The replacement evaluation holds every input tensor at the intact length of 2,295 tokens by adding filler tokens after all scored targets. Deletion still shifts the real choice tokens; suffix padding only fixes the numerical tensor shape. An uncached causal forward pass prevents the future filler from entering any scored history. Across all 1,750 exports, constant replacement, placeholders, and deletion have exactly zero first-choice discrepancy relative to the original evaluation. Same-shape future-token probes also give zero discrepancy. The history rewrite changes the initial instruction, so first-choice equality with the original prompt is not required for that operation.

Independent-test exports pass participant, action, donor-map, checkpoint-manifest, and probability checks. Three restarted restless jobs retained the successful zero-discrepancy causal probes from their original logs; log hashes are archived because resumed completion records omit probes for already saved participants. Test spatial exports use the fixed-size batches and causal-prefix checks described below.

Oracle replay is checked against the probabilities saved during generation. Participant identifiers, actual choices, trial windows, and donor assignments are aligned before comparisons. Donor maps are full derangements, probabilities are finite and normalized, and the difference between donor and intact NLL matches the independently computed effect. At zero spatial reward weight, the oracle's intact and donor policies agree exactly.

For neural evaluations, causal-prefix invariance is checked under reward replacement: changing a later reward cannot affect a preceding choice probability. This check matters with quantized, accelerated inference. An earlier variable-padding spatial LLaMA evaluation changed first-choice probabilities despite identical causal inputs, reaching an absolute discrepancy of approximately $.072$. Those results are excluded. The included reevaluation uses fixed batch size four and padded length 1,024 with uncached Unsloth forward passes. Across the actual intact/donor comparisons, the first-choice discrepancy is exactly zero in the final exports. Audits cover seven weights, both input conditions, and all 50 participants. Numerical stability is therefore checked on the operation being interpreted, not inferred from successful training alone.

Intervals use 2,000 participant resamples, averaging donor repetitions and scratch-network seed metrics first. They condition on fitted checkpoints and donor maps, not all training uncertainty. Comparisons are not multiplicity-adjusted or preregistered. Main hypotheses were developed on validation results and evaluated on independent test participants without checkpoint reselection. Supplementary operation and neural-generator controls remain development analyses.


\begin{thebibliography}{22}
\providecommand{\natexlab}[1]{#1}
\providecommand{\url}[1]{\texttt{#1}}
\expandafter\ifx\csname urlstyle\endcsname\relax
  \providecommand{\doi}[1]{doi: #1}\else
  \providecommand{\doi}{doi: \begingroup \urlstyle{rm}\Url}\fi

\bibitem[Adebayo et~al.(2018)Adebayo, Gilmer, Muelly, Goodfellow, Hardt, and
  Kim]{adebayo2018sanity}
Julius Adebayo, Justin Gilmer, Michael Muelly, Ian Goodfellow, Moritz Hardt,
  and Been Kim.
\newblock Sanity checks for saliency maps.
\newblock In S.~Bengio, H.~Wallach, H.~Larochelle, K.~Grauman, N.~Cesa-Bianchi,
  and R.~Garnett (eds.), \emph{Advances in Neural Information Processing
  Systems}, volume~31. Curran Associates, Inc., 2018.
\newblock URL
  \url{https://proceedings.neurips.cc/paper_files/paper/2018/file/294a8ed24b1ad22ec2e7efea049b8737-Paper.pdf}.

\bibitem[Binz et~al.(2025)Binz, Akata, Bethge, Br{\"a}ndle, Callaway,
  Coda-Forno, Dayan, Demircan, Eckstein, {\'E}ltet{\H{o}},
  et~al.]{binz2025foundation}
Marcel Binz, Elif Akata, Matthias Bethge, Franziska Br{\"a}ndle, Fred Callaway,
  Julian Coda-Forno, Peter Dayan, Can Demircan, Maria~K Eckstein, No{\'e}mi
  {\'E}ltet{\H{o}}, et~al.
\newblock {A foundation model to predict and capture human cognition}.
\newblock \emph{Nature}, 644\penalty0 (8078):\penalty0 1002--1009, 2025.
\newblock \doi{10.1038/s41586-025-09215-4}.
\newblock URL \url{https://doi.org/10.1038/s41586-025-09215-4}.

\bibitem[Cho et~al.(2014)Cho, van Merri{\"e}nboer, Gulcehre, Bahdanau,
  Bougares, Schwenk, and Bengio]{cho2014learning}
Kyunghyun Cho, Bart van Merri{\"e}nboer, Caglar Gulcehre, Dzmitry Bahdanau,
  Fethi Bougares, Holger Schwenk, and Yoshua Bengio.
\newblock {Learning phrase representations using RNN encoder--decoder for
  statistical machine translation}.
\newblock In \emph{Proceedings of the 2014 Conference on Empirical Methods in
  Natural Language Processing ({EMNLP})}, pp.\  1724--1734. Association for
  Computational Linguistics, 2014.
\newblock \doi{10.3115/v1/D14-1179}.
\newblock URL \url{https://aclanthology.org/D14-1179/}.

\bibitem[Covert et~al.(2021)Covert, Lundberg, and Lee]{covert2021explaining}
Ian Covert, Scott Lundberg, and Su-In Lee.
\newblock {Explaining by removing: A unified framework for model explanation}.
\newblock \emph{Journal of Machine Learning Research}, 22\penalty0
  (209):\penalty0 1--90, 2021.
\newblock URL \url{https://www.jmlr.org/papers/v22/20-1316.html}.

\bibitem[Daw \& Doya(2006)Daw and Doya]{daw2006computational}
Nathaniel~D. Daw and Kenji Doya.
\newblock The computational neurobiology of learning and reward.
\newblock \emph{Current Opinion in Neurobiology}, 16\penalty0 (2):\penalty0
  199--204, 2006.
\newblock \doi{10.1016/j.conb.2006.03.006}.
\newblock URL \url{https://doi.org/10.1016/j.conb.2006.03.006}.

\bibitem[Dettmers et~al.(2023)Dettmers, Pagnoni, Holtzman, and
  Zettlemoyer]{dettmers2023qlora}
Tim Dettmers, Artidoro Pagnoni, Ari Holtzman, and Luke Zettlemoyer.
\newblock {QLoRA}: Efficient finetuning of quantized {LLMs}.
\newblock In A.~Oh, T.~Naumann, A.~Globerson, K.~Saenko, M.~Hardt, and
  S.~Levine (eds.), \emph{Advances in Neural Information Processing Systems},
  volume~36, pp.\  10088--10115. Curran Associates, Inc., 2023.
\newblock \doi{10.52202/075280-0441}.
\newblock URL
  \url{https://proceedings.neurips.cc/paper_files/paper/2023/file/1feb87871436031bdc0f2beaa62a049b-Paper-Conference.pdf}.

\bibitem[Dezfouli et~al.(2019)Dezfouli, Griffiths, Ramos, Dayan, and
  Balleine]{dezfouli2019models}
Amir Dezfouli, Kristi Griffiths, Fabio Ramos, Peter Dayan, and Bernard~W
  Balleine.
\newblock {Models that learn how humans learn: The case of decision-making and
  its disorders}.
\newblock \emph{PLOS Computational Biology}, 15\penalty0 (6):\penalty0
  e1006903, 2019.
\newblock \doi{10.1371/journal.pcbi.1006903}.
\newblock URL \url{https://doi.org/10.1371/journal.pcbi.1006903}.

\bibitem[Eckstein et~al.(2026)Eckstein, Summerfield, Daw, and
  Miller]{eckstein2026hybrid}
Maria~K. Eckstein, Christopher Summerfield, Nathaniel~D. Daw, and Kevin~J.
  Miller.
\newblock Hybrid neural--cognitive models reveal how memory shapes human reward
  learning.
\newblock \emph{Nature Human Behaviour}, 10\penalty0 (5):\penalty0 972--987,
  2026.
\newblock \doi{10.1038/s41562-025-02324-0}.
\newblock URL \url{https://doi.org/10.1038/s41562-025-02324-0}.

\bibitem[Geirhos et~al.(2020)Geirhos, Jacobsen, Michaelis, Zemel, Brendel,
  Bethge, and Wichmann]{geirhos2020shortcut}
Robert Geirhos, J{\"o}rn-Henrik Jacobsen, Claudio Michaelis, Richard Zemel,
  Wieland Brendel, Matthias Bethge, and Felix~A Wichmann.
\newblock {Shortcut learning in deep neural networks}.
\newblock \emph{Nature Machine Intelligence}, 2\penalty0 (11):\penalty0
  665--673, 2020.
\newblock \doi{10.1038/s42256-020-00257-z}.
\newblock URL \url{https://doi.org/10.1038/s42256-020-00257-z}.

\bibitem[Gershman(2020)]{gershman2020origin}
Samuel~J. Gershman.
\newblock Origin of perseveration in the trade-off between reward and
  complexity.
\newblock \emph{Cognition}, 204:\penalty0 104394, 2020.
\newblock \doi{10.1016/j.cognition.2020.104394}.
\newblock URL \url{https://doi.org/10.1016/j.cognition.2020.104394}.

\bibitem[Grattafiori et~al.(2024)Grattafiori, Dubey, Jauhri, Pandey, Kadian,
  Al-Dahle, Letman, Mathur, Schelten, Vaughan, et~al.]{grattafiori2024llama}
Aaron Grattafiori, Abhimanyu Dubey, Abhinav Jauhri, Abhinav Pandey, Abhishek
  Kadian, Ahmad Al-Dahle, Aiesha Letman, Akhil Mathur, Alan Schelten, Alex
  Vaughan, et~al.
\newblock {The Llama 3 Herd of Models}.
\newblock \emph{arXiv preprint arXiv:2407.21783}, 2024.
\newblock URL \url{https://arxiv.org/abs/2407.21783}.

\bibitem[Hooker et~al.(2019)Hooker, Erhan, Kindermans, and
  Kim]{NEURIPS2019_fe4b8556}
Sara Hooker, Dumitru Erhan, Pieter-Jan Kindermans, and Been Kim.
\newblock A benchmark for interpretability methods in deep neural networks.
\newblock In H.~Wallach, H.~Larochelle, A.~Beygelzimer, F.~d\textquotesingle
  Alch\'{e}-Buc, E.~Fox, and R.~Garnett (eds.), \emph{Advances in Neural
  Information Processing Systems}, volume~32. Curran Associates, Inc., 2019.
\newblock URL
  \url{https://proceedings.neurips.cc/paper_files/paper/2019/file/fe4b8556000d0f0cae99daa5c5c5a410-Paper.pdf}.

\bibitem[Hu et~al.(2021)Hu, Shen, Wallis, Allen-Zhu, Li, Wang, Wang, and
  Chen]{hu2021lora}
Edward~J Hu, Yelong Shen, Phillip Wallis, Zeyuan Allen-Zhu, Yuanzhi Li, Shean
  Wang, Lu~Wang, and Weizhu Chen.
\newblock {LoRA: Low-Rank Adaptation of Large Language Models}.
\newblock \emph{arXiv preprint arXiv:2106.09685}, 2021.
\newblock URL \url{https://arxiv.org/abs/2106.09685}.

\bibitem[Ji-An et~al.(2025)Ji-An, Benna, and Mattar]{ji2025discovering}
Li~Ji-An, Marcus~K Benna, and Marcelo~G Mattar.
\newblock {Discovering cognitive strategies with tiny recurrent neural
  networks}.
\newblock \emph{Nature}, 644\penalty0 (8078):\penalty0 993--1001, 2025.
\newblock \doi{10.1038/s41586-025-09142-4}.
\newblock URL \url{https://doi.org/10.1038/s41586-025-09142-4}.

\bibitem[Lau \& Glimcher(2005)Lau and Glimcher]{lau2005dynamic}
Brian Lau and Paul~W Glimcher.
\newblock {Dynamic response-by-response models of matching behavior in rhesus
  monkeys}.
\newblock \emph{Journal of the Experimental Analysis of Behavior}, 84\penalty0
  (3):\penalty0 555--579, 2005.
\newblock \doi{10.1901/jeab.2005.110-04}.
\newblock URL \url{https://doi.org/10.1901/jeab.2005.110-04}.

\bibitem[Oh \& Gobet(2026)Oh and Gobet]{oh2026smallcogfm}
Nick Oh and Fernand Gobet.
\newblock Small foundation models of human cognition and behaviour.
\newblock In \emph{Third Conference on Language Modeling (COLM)}, 2026.
\newblock URL \url{https://arxiv.org/abs/2608.05224v3}.
\newblock arXiv:2608.05224.

\bibitem[Peterson et~al.(2021)Peterson, Bourgin, Agrawal, Reichman, and
  Griffiths]{peterson2021using}
Joshua~C Peterson, David~D Bourgin, Mayank Agrawal, Daniel Reichman, and
  Thomas~L Griffiths.
\newblock {Using large-scale experiments and machine learning to discover
  theories of human decision-making}.
\newblock \emph{Science}, 372\penalty0 (6547):\penalty0 1209--1214, 2021.
\newblock \doi{10.1126/science.abe2629}.
\newblock URL \url{https://doi.org/10.1126/science.abe2629}.

\bibitem[Schultz et~al.(1997)Schultz, Dayan, and Montague]{schultz1997neural}
Wolfram Schultz, Peter Dayan, and P.~Read Montague.
\newblock A neural substrate of prediction and reward.
\newblock \emph{Science}, 275\penalty0 (5306):\penalty0 1593--1599, 1997.
\newblock \doi{10.1126/science.275.5306.1593}.
\newblock URL \url{https://doi.org/10.1126/science.275.5306.1593}.

\bibitem[Vaswani et~al.(2017)Vaswani, Shazeer, Parmar, Uszkoreit, Jones, Gomez,
  Kaiser, and Polosukhin]{vaswani2017attention}
Ashish Vaswani, Noam Shazeer, Niki Parmar, Jakob Uszkoreit, Llion Jones,
  Aidan~N Gomez, {\L}ukasz Kaiser, and Illia Polosukhin.
\newblock Attention is all you need.
\newblock In I.~Guyon, U.~Von Luxburg, S.~Bengio, H.~Wallach, R.~Fergus,
  S.~Vishwanathan, and R.~Garnett (eds.), \emph{Advances in Neural Information
  Processing Systems}, volume~30. Curran Associates, Inc., 2017.
\newblock URL
  \url{https://proceedings.neurips.cc/paper_files/paper/2017/file/3f5ee243547dee91fbd053c1c4a845aa-Paper.pdf}.

\bibitem[Wilson \& Collins(2019)Wilson and Collins]{wilson2019ten}
Robert~C Wilson and Anne~GE Collins.
\newblock {Ten simple rules for the computational modeling of behavioral data}.
\newblock \emph{eLife}, 8:\penalty0 e49547, 2019.
\newblock \doi{10.7554/eLife.49547}.
\newblock URL \url{https://doi.org/10.7554/eLife.49547}.

\bibitem[Wu et~al.(2018)Wu, Schulz, Speekenbrink, Nelson, and
  Meder]{wu2018generalization}
Charley~M Wu, Eric Schulz, Maarten Speekenbrink, Jonathan~D Nelson, and
  Bj{\"o}rn Meder.
\newblock {Generalization guides human exploration in vast decision spaces}.
\newblock \emph{Nature Human Behaviour}, 2\penalty0 (12):\penalty0 915--924,
  2018.
\newblock \doi{10.1038/s41562-018-0467-4}.
\newblock URL \url{https://doi.org/10.1038/s41562-018-0467-4}.

\bibitem[Xie \& Zhu(2025)Xie and Zhu]{xie2025centaur}
Hanbo Xie and Jian-Qiao Zhu.
\newblock {Centaur may have learned a shortcut that explains away psychological
  tasks}.
\newblock PsyArXiv preprint, 2025.
\newblock URL \url{https://doi.org/10.31234/osf.io/u7z4t_v1}.

\end{thebibliography}
\end{document}